\documentstyle[twocolumn,aps]{revtex}

\begin{document}
\draft

\twocolumn[\hsize\textwidth\columnwidth\hsize\csname @twocolumnfalse\endcsname

\title{One-particle interchain hopping in coupled Hubbard chains}

\author{D. Poilblanc$^{1}$, H. Endres$^{2}$, F. Mila$^{1}$,
M. G. Zacher$^{2}$, S. Capponi$^{1}$ and W. Hanke$^{2}$}
\address{
$^{1}$Laboratoire de Physique Quantique, Universit\'e Paul Sabatier,
31062 Toulouse, France\\
$^{2}$Institute for Physics, University of Wurzburg, D-97074 Wurzburg, Germany\\
}

\date{October 95} 
\maketitle 

\begin{abstract}
Interchain hopping in systems of coupled chains of correlated electrons 
is investigated by exact diagonalizations
and Quantum-Monte-Carlo methods. For two weakly coupled Hubbard chains
at commensurate densities (e.g. n=1/3) the splitting at the Fermi level 
between bonding and antibonding bands is strongly reduced (but not suppressed)
by repulsive interactions extending to a few lattice spacings. The magnitude
of this reduction is directly connected to the exponent 
$\alpha$ of the 1D Luttinger liquid. However, we show that 
the incoherent part of the single particle spectral function is much less
affected by the interchain coupling. This suggests that incoherent 
interchain hopping could occur for intermediate $\alpha$ values.

\end{abstract}

\pacs{ PACS Numbers: 71.27.+a, 71.38.+i, 74.20.Mn, 74.25.Kc}
\vskip2pc]
\narrowtext

The possible occurence of incoherent single-particle hopping in
strongly correlated systems is presently actively debated.
Such an unusual behavior could, for example, explain the unconventional c-axis
conductivity found in two-layered high-$T_c$ compounds \cite{c_transport}.
In addition, the high transition temperatures of these materials
may also be connected to incoherent hopping \cite{anderson}:
as a consequence of incoherence, the coupled planes can lower
the total energy by coherent pair-hopping processes, leading to an 
amplification of the tendency towards superconducting pairing. 

Perturbation and renormalisation-group treatments
\cite{schulz,fabrizio,bourbonnais} of the transverse hopping between
chains reveals that the key parameter which governs interchain hopping
is the non-universal exponent $\alpha$ of the 1D Luttinger liquid (LL)
\cite{haldane} characterizing the low frequency behavior of the density 
of state
$N(\omega)\sim \omega^\alpha$. For $\alpha\ge 1$, ie for sufficiently
strong interaction, the interchain hopping becomes irrelevant.
Recently Clarke, Strong and Anderson \cite{clarke} suggested a somewhat 
different criterium: let us consider, for (real) times $\tau <0$,
a system of two decoupled chains with different electron numbers on each
chain in some initial state $|\Psi_0\big>$.
Then, if the interchain hopping $t_\perp$ is switched on at $\tau=0$
two different behaviors can be observed: the quantity
$P(\tau)=|A(\tau)|^2$, with $A(\tau)=\big<\Psi_0|e^{iH\tau}|\Psi_0\big>$ 
corresponding to the probability 
for the system to be in its initial state, at a later time $\tau > 0$ might
show oscillatory or monotoneous behaviors characterizing
``coherent'' or ``incoherent'' interchain hopping, respectively. 
It was suggested that such an incoherent behavior could occur at 
values of $\alpha$ as small as $0.5$.

The first numerical attempt to examine the question of coherent versus
incoherent single particle hopping was initiated by two of us \cite{MP}. 
By exact diagonalizations of two coupled t-J chains, 
an interesting connection between coherence and integrability was found. 
However, it should be emphasized that, in this approach, the initial state
$|\Psi_0\big>$ at $\tau=0$ is far from equilibrium. Indeed, constructing the
whole system with $t_\perp=0$ corresponds to a macroscopic 
perturbation for the full hamiltonian (including the tranverse hopping).
Interchain {\it single} particle hopping in the true GS ($t_\perp\ne 0$)
is still an open question. 

For this purpose, let us first consider two weakly coupled 
chains described by an extended Hubbard model,
\begin{eqnarray}
\label{hami}
H &=&-t\sum_{i,\lambda,\sigma}
(c_{i,\lambda,\sigma}^\dagger c_{i+1,\lambda,\sigma} + h.c.)
+ U\sum_{i,\lambda} n_{i,\lambda,\uparrow} n_{i,\lambda,\downarrow}
\nonumber  
\\
&+& \sum_{r,i,\lambda} V_r n_{i,\lambda} n_{i+r,\lambda}
- t_\perp \sum_{i,\lambda, \sigma}
            ( c_{i,1,\sigma}^\dagger c_{i,2,\sigma} + h.c.),
\end{eqnarray}
\noindent
where $c_{i,\lambda,\sigma}^\dagger$ is a creation fermion operator
at site i of the chain $\lambda$ (=1,2). 
We shall deal here with two cases; (i) a purely on-site interaction
($V_r=0$) or (ii) an extended repulsion of the form $V_r=U/(r+1)$ 
limited to r=1 and 2. 

In contrast to previous numerical work, we shall assume that,
at $\tau<0$, the system contains $M$ electrons in the ground state (GS)
$|\phi_0\big>$ of the full hamiltonian (\ref{hami}) and that, at $\tau=0$,
only a {\it single} extra 
particle (e.g. a hole) of longitudinal momentum k is added to, let say, 
chain number 1.  Provided the occupation number 
$n_{k,1}=\big<\phi_0|c_{k,1,\sigma}^\dagger c_{k,1,\sigma}|\phi_0\big>$
is non zero, the corresponding probability $P(\tau)$ 
is then defined by the amplitude 
$A(\tau)=\int_{-\infty}^\infty A(\omega)
e^{i\omega\tau} d\tau\, / \,n_{k,1}$, with 
\begin{eqnarray}
\label{spec}
A(\omega)=-\frac{1}{\pi}Im\{\big<\phi_0| c_{k,1,\sigma}^\dagger
\frac{1}{\omega +i\epsilon-H+E_0^M}c_{k,1,\sigma}|\phi_0\big>\}.
\end{eqnarray}
\noindent
This corresponds to the particular choice of an initial (normalized)
state of the form
$|\Psi_0\big>=\frac{1}{(n_{k,1})^{1/2}}c_{k,1,\sigma}|\phi_0\big>$. 
Note that when an electron is added instead of a hole 
the creation and annihilation operators
in (\ref{spec}) have to be interchanged. At finite temperature $T=1/\beta$
all expectation values in $|\phi_0\big>$ must be replaced by thermal averages.  

Writing $c_{k,1,\sigma}$ in terms of the bonding and anti-bonding 
operators $c_{k,k_\perp,\sigma}$ ($k_\perp=0$, $\pi$) 
it is straightforward to express (\ref{spec}) as a linear combination,
$A(\omega)=(A(k,0,\omega)+A(k,\pi,\omega))
/(1+e^{\beta(\omega-\mu)})$ where 
$A({\bf k},\omega)=-\frac{1}{\pi}Im\,G({\bf k},\omega)$ corresponds 
to the usual spectral function in momentum space (${\bf k}=(k,k_\perp)$). 
At this point, it is instructive to consider the non-interacting U=0
limit (and $T=0$). In this case, 
$A(\bf k,\omega)\sim\delta (\omega-\epsilon_{\bf k})$ and 
the two bands $\epsilon_{\bf k}$ for $k_\perp=0$ and $\pi$ are split 
by $2t_\perp$. If the momentum k (close to $k_F$) is chosen in such a 
way that both $k_\perp=0$ and $k_\perp=\pi$ states in $|\phi_0\big>$ are
occupied, $P(\tau)$ acquires an oscillatory behavior with a characteristic 
time scale $\pi/t_{\perp}$. 

For a generic LL the quasiparticle 
$\delta$-peak is in fact replaced by a 
singularity \cite{spec_func} corresponding to a 
collective excitation (in fact split holon and spinon peaks). 
However, it is clear that an oscillatory 
behavior of $P(\tau)$
is only possible if $t_\perp$ splits the dispersion relation of the bonding
and anti-bonding modes by a finite amount although this alone
might not be sufficient. 
One possible scenario would be that the whole spectra 
for $k_\perp=0$ or $\pi$ (i.e. the poles of the
one particle Green functions) would be displaced by $\pm t_\perp$ respectively.
In this case, $P(\tau)$ would exactly take the form
$\cos^2(t_\perp\tau)P_{1D}(\tau)$ where $P_{1D}(\tau)$ is independent of
$t_\perp$ and is directly related to the 1D spectral function
$A_{1D}(k,\omega)$ of the LL. However, we expect that, 
in general, (i) the incoherent part of $A_{1D}(k,\omega)$ (i.e the part 
not related to the collective mode) is less 
sensitive to the transverse hopping at energies $\omega\gg t_\perp$ 
and (ii) the relative weight of 
this part increases with the strength and the range of the interaction
within the chains. 
Since analytic approaches of such a problem are 
extremely difficult, we shall rely in the following on exact diagonalizations
of finite clusters (where dynamical and time dependent quantities can be
calculated exactly) and on Quantum Monte Carlo (QMC) data. 

As a preliminary study, the various LL parameters of the 1D 
{\it single chain} (e.g. $\alpha$, $K_\rho$,
spin and charge velocities, etc...) are determined.
The parameters of the 1D on-site Hubbard model are known from 
numerical calculations based on the Bethe Ansatz \cite{schulz1}. 
In the case of the extended model these parameters are obtained 
by standard finite-size scaling of ED 
results. For more technical details the reader can refer to Ref. \cite{mila}.
Hereafter, we shall restrict ourselves to a fermion density of n=1/3 where
relatively large values of the exponent $\alpha$ (up to $\sim 1.78$) 
can in principal be obtained \cite{schulz2}. Our numerical data show that 
the maximum value can be realized in the case of the extended model 
for $U\sim 10.5$. Above this value a gap appears in the single particle 
spectral density possibly signaling a metal-insulator transition of no
interest for us in the present work. 

The result for the spectral function of the double Hubbard chain deduced
by analytic continuation of the QMC data by maximum entropy methods 
is shown in Fig. (\ref{fig1}) for $U/t=8$. Bonding and anti-bonding bands 
dispersing through the fermi level ($\omega=\mu=0$) can be clearly seen. 
Although error bars are rather large \cite{note} a splitting $\Delta E$
of order $2t_\perp$ 
can be measured. It is {\it a priori} puzzling that such a large U does
not lead to a substantial reduction of the splitting. However,
one has to remember that $\alpha$ remains very small for an on-site
interaction, e.g. $\alpha\sim 0.07$ for $U/t=8$.

As a comparison, the dispersions of the two low-energy branches 
corresponding to sharp peaks in $A(\bf k,\omega)$ for 
$k_\perp=0$ and  $k_\perp=\pi$ and obtained by ED techniques are shown in 
Fig. (\ref{fig2}) for the same value of U, in qualitative agreement 
with the QMC results. Note that the boundary conditions along the chain
direction (periodic or anti-periodic) are always chosen in such a way that the
GS $|\phi_0\big>$ has a closed shell configuration.
It should be emphasized that the band splitting picture is only valid 
sufficiently close to the Fermi level $\mu$ and that, on finite clusters,
different values are obtained immediately above (i.e. $|k|>k_F$) or below 
(i.e $|k|<k_F$) the chemical potentail $\mu$. 

A careful study of the ED data reveals that $\Delta E$, for sufficiently 
small $t_\perp$, behaves like $a t_\perp + b t_\perp^3$, where $a$ and $b$ 
depend on the parameters and system size. The ratio 
in the limit of vanishing $t_\perp$ is then easily extracted and the results 
are displayed in Fig. (\ref{fig3}a) for the Hubbard and the extended
Hubbard models and various system size. 
Assuming finite size corrections of the form $1/L_x$
\cite{note2}, where $L_x$ is the length of the chains, and considering the 
average between the splittings above and below $\mu$, one can also estimate 
$\Delta E/2t_\perp$ in the thermodynamic limit as seen in Fig. (\ref{fig3}b).
Clearly repulsion at intermediate distances is much more effective to 
reduce the splitting. However, the plot of the same data
vs $\alpha$ instead of U in (\ref{fig3}b) suggests that 
$\alpha$ alone seems to be the key parameter controling the splitting.
Indeed, an extended interaction leads to larger $\alpha$ values and hence to
smaller $\Delta E/2t_\perp$. 

The previous study suggests that, at least for $\alpha<1$,
a tranverse hopping $t_\perp$ leads to a splitting of the bonding
and anti-bonding branches in agreement with perturbative treatments
of $t_\perp$ \cite{schulz,fabrizio,bourbonnais}. However, the expression
for $P(\tau)$ involves the whole spectral function $A(\bf k,\omega)$ at
all frequencies so that it is not clear whether a spliting of the 
low energy singularity alone will necessarily lead to a coherent 
interchain behavior. The spectral functions at $k_\perp=0$ and
$k_\perp=\pi$ of the on-site Hubbard model and for a momentum k close to $k_F$
are shown in Fig. (\ref{fig4}a). Most of the spectral weight is located
at the low energy singularity which is clearly split by the transverse 
hopping $t_\perp$. On the contrary, in the case of the extended Hubbard
model (with parameters such that $\alpha\sim 0.53$) shown 
in Fig. (\ref{fig4}b), more spectral weight appears 
far from the Fermi level and is less affected by $t_\perp$. 

The probability $P(\tau)$ has been computed from the knowledge of
the spectral functions shown in Figs. (\ref{fig4}a,b) 
and results are shown in Fig. (\ref{fig5}). For the two types of models 
we observe both fast and slow oscillations which have different 
physical origins. The fast oscillations are due to the fine structure 
of the 1D spectral function $A_{1D}(\bf k,\omega)$ (i.e. for $t_\perp=0$);
for example, the sharp peaks of Fig. (\ref{fig4}b) might be attributed 
to shadow bands \cite{mila2}. We expect that, in the thermodynamic limit, 
these peaks will broaden and hence the fast oscillations will get
strongly damped if not suppressed. More interestingly, the slow oscillations
have a pseudo-periodicity of order $\pi/\Delta E$ and are due to the
effect of $t_\perp$. They reveal some degree of interchain coherence. 
However, in the case of the extended Hubbard model with $\alpha\sim 0.53$ 
a larger damping can be seen. It is not clear yet whether these slow 
oscillations will completely desappear in the thermodynamic limit. 

The previous results suggest that a splitting of the dispersion of the 
single-particle low-energy mode of the LL by the tranverse coupling alone 
does not necessarily lead to an oscillatory behavior of $P(\tau)$ 
which, according to Ref. \cite{clarke}, characterizes a coherent interchain
hopping. Indeed, for a sufficiently repulsive intermediate range (e.g. up to
distances of 2 or 3 lattice spacings) interaction the large incoherent 
part of the single-particle 1D spectral function might play a very important 
role in preventing coherent interchain hopping. 

DP acknowledges support from the EEC Human Capital and Mobility 
program under Grant No. CHRX-CT930332.
WH acknowledges EC Grant No. ERBCHRX-CT940438 and the Bavarian FORSUPRA Progam.
We also thank IDRIS, Orsay (France), 
for allocation of CPU time on the C94 and C98 CRAY supercomputers,
and the HLRZ J\"ulich and LRZ M\"unchen for access to the 
CRAY Y-MP supercomputers.

%
%
\begin{figure}
\caption{ QMC result for $A({\bf k},\omega)$ vs ${\bf k}=(k,k_\perp)$
on a $2\times 24$ Hubbard model with $U/t=8$, $t_\perp/t=0.15$  
at $n=1/3$ and an inverse temperature of $\beta t=10$.
Part (a) and (b) are density plots for the $k_\perp =0$ and $k_\perp =\pi$ 
branches respectively with the darker shading corresponding to higher
spectral weight, and the points with error bars showing the position 
of the peaks.
\label{fig1}
}
\end{figure}

%
%
\begin{figure}
\caption{ Exact diagonalization result for the lowest excitation energies
vs ${\bf k}=(k,k_\perp)$ obtained on a $2\times 12$ Hubbard model 
with $U/t=8$, $t_\perp/t=0.15$ at $n=1/3$.
The full symbols correspond to the main bonding and antibonding branches.
Other low energy poles in $A_{{\bf k},\omega}$ (with smaller spectral weight)
are also indicated by open dots. The thin dotted lines correspond to 
the $U=0$ case and the dashed line to the chemical potential. 
\label{fig2}
}
\end{figure}

%
%
\begin{figure}
\caption{ (a) Normalized splitting $\Delta E/2t_\perp$ extrapolated to 
$t_\perp\rightarrow 0$ (see text) vs U
obtained by exact diagonalizations of $2\times 6$ (open symbols) 
and $2\times 12$ (full symbols) systems with on-site (O-S) Hubbard
or intermediate range (I-R) repulsive interactions (see text).
Both momenta $k>k_F$ (+) and $k<k_F$ (-) in the vicinity of $k_F$ have been 
considered as indicated on the plot; (b) same quantity plotted as a function 
of $\alpha$. The points with error bars correspond to crude estimates 
for the thermodynamic limit $N\rightarrow\infty$ (see text). 
Error bars are estimated from separate extrapolations of the (+) and (-) data.
\label{fig3}
}
\end{figure}

%
%
\begin{figure}
\caption{ Spectral function $A(k,k_\perp,\omega)$ vs $\omega$ on a 
$2\times 12$ system with $n=1/3$ filling and for a 
longitudinal momentum $k=\pi/12 < k_F$. The full and dashed lines correspond to
$k_\perp=\pi$ and $k_\perp=0$, respectively. Only the $\omega < \mu$ region
is shown (occupied states). (a) and (b) correspond to an 
on-site $U=8$ Hubbard repulsion and to an intermediate range repulsion,
$U=6$, $V_1=3$ and $V_2=2$, respectively.
\label{fig4}
}
\end{figure}

%
%
\begin{figure}
\caption{Probability $P(\tau)$ for a 
$2\times 12$ system at $n=1/3$ filling and an extra hole
with a longitudinal momentum $k=\pi/12 < k_F$. 
The dashed and full lines correspond to an 
on-site Hubbard repulsion $U=8$ and to an intermediate range repulsion,
with $U=6$, $V_1=3$ and $V_2=2$, respectively.
\label{fig5}
}
\end{figure}

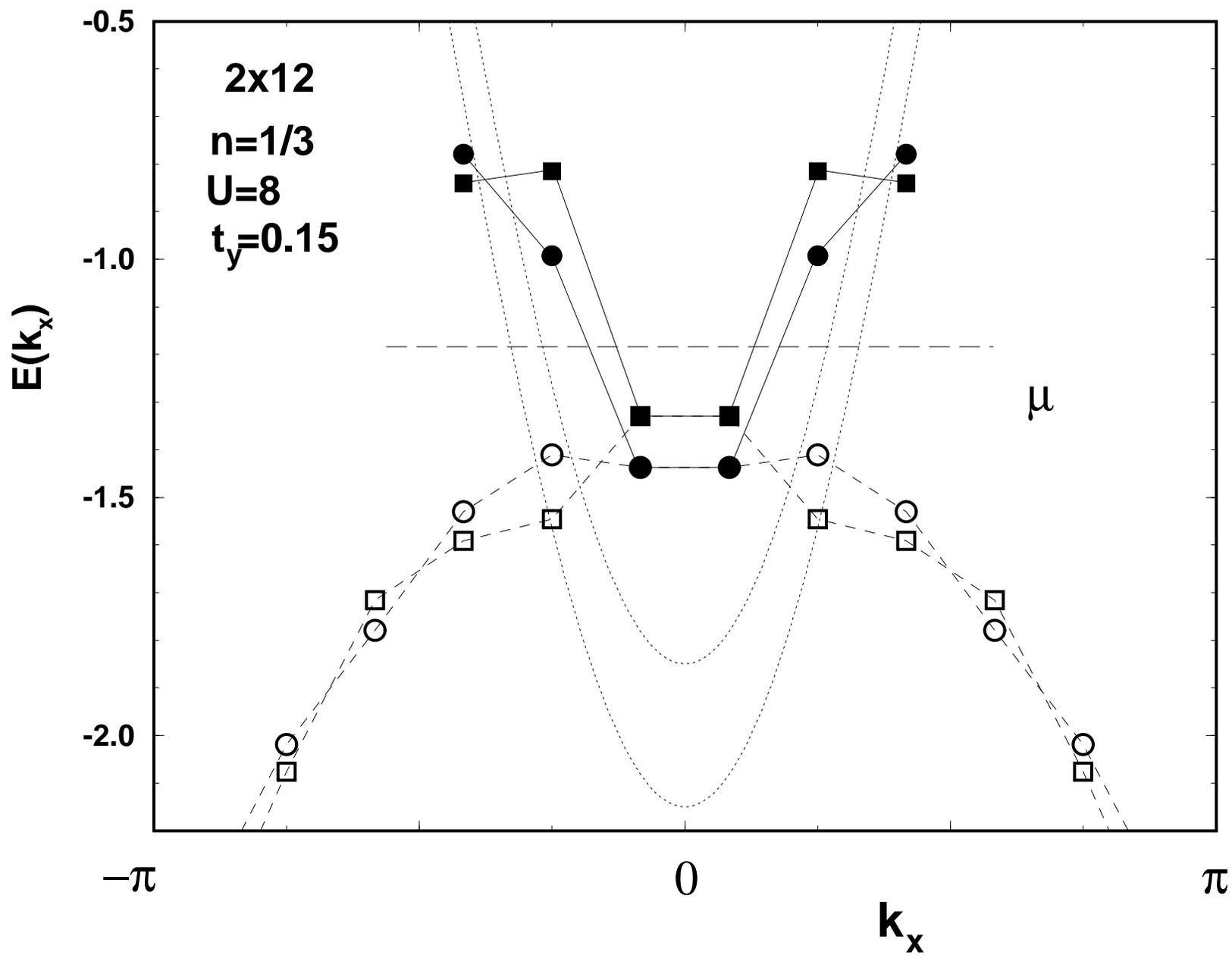
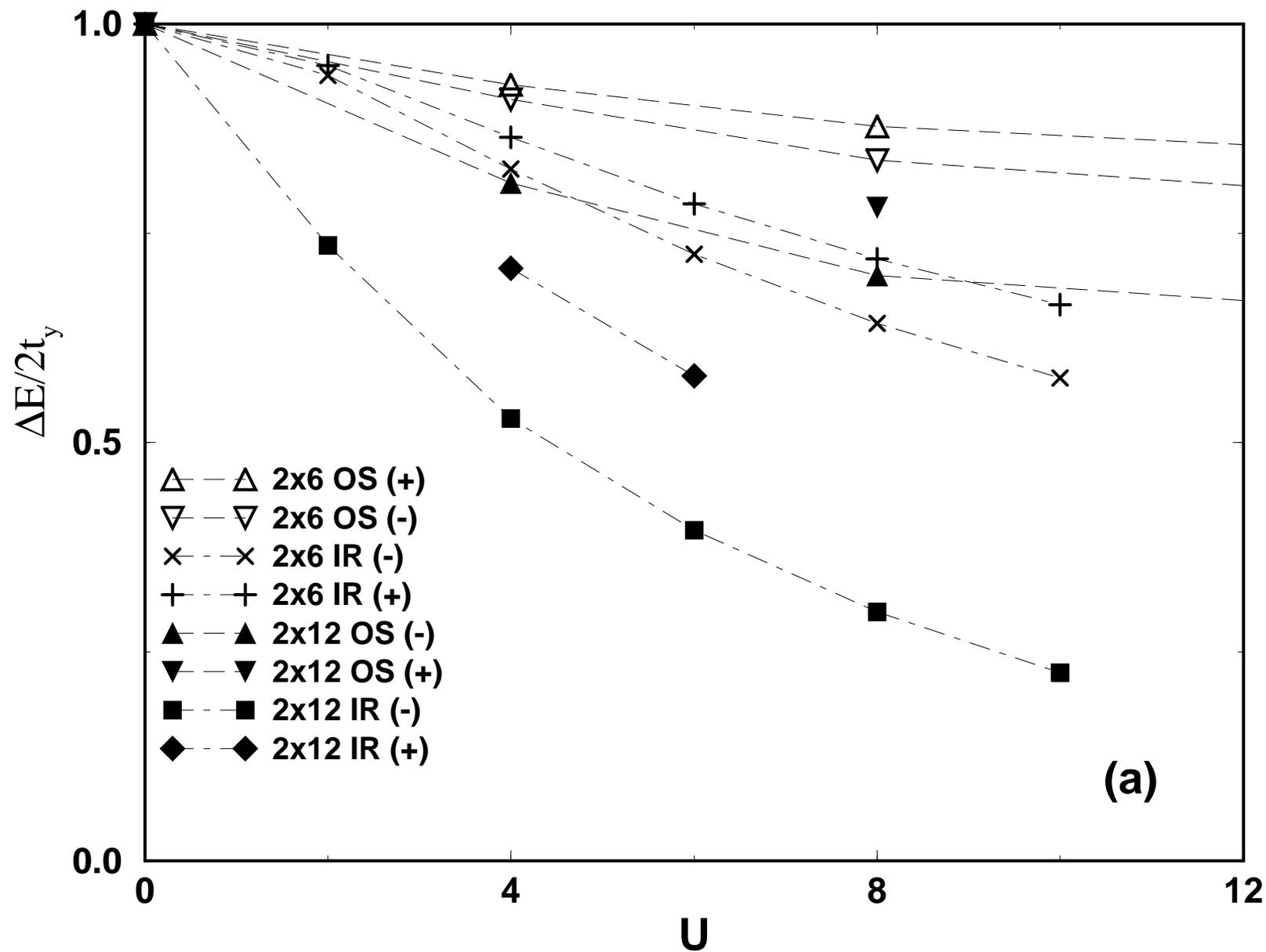
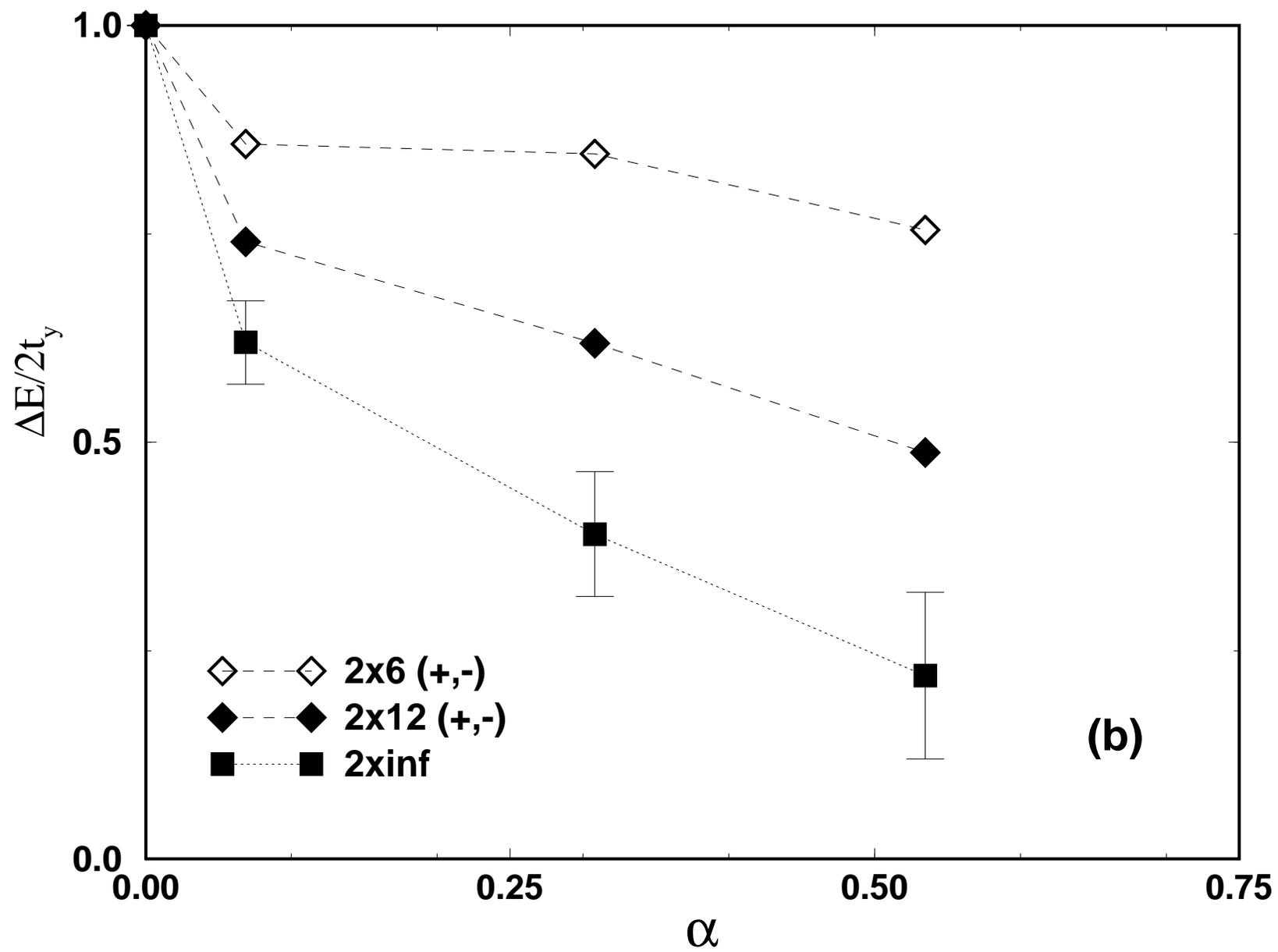
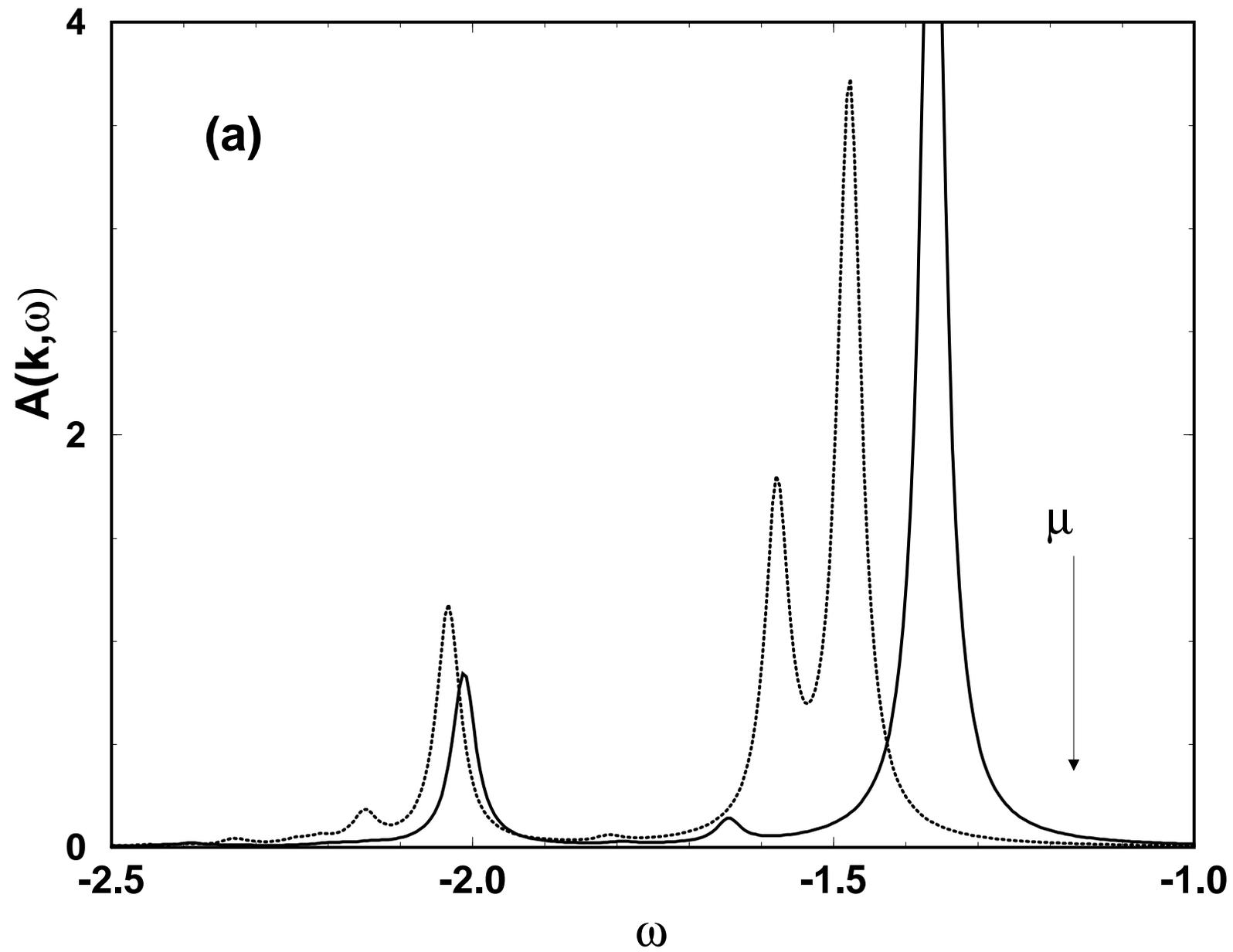
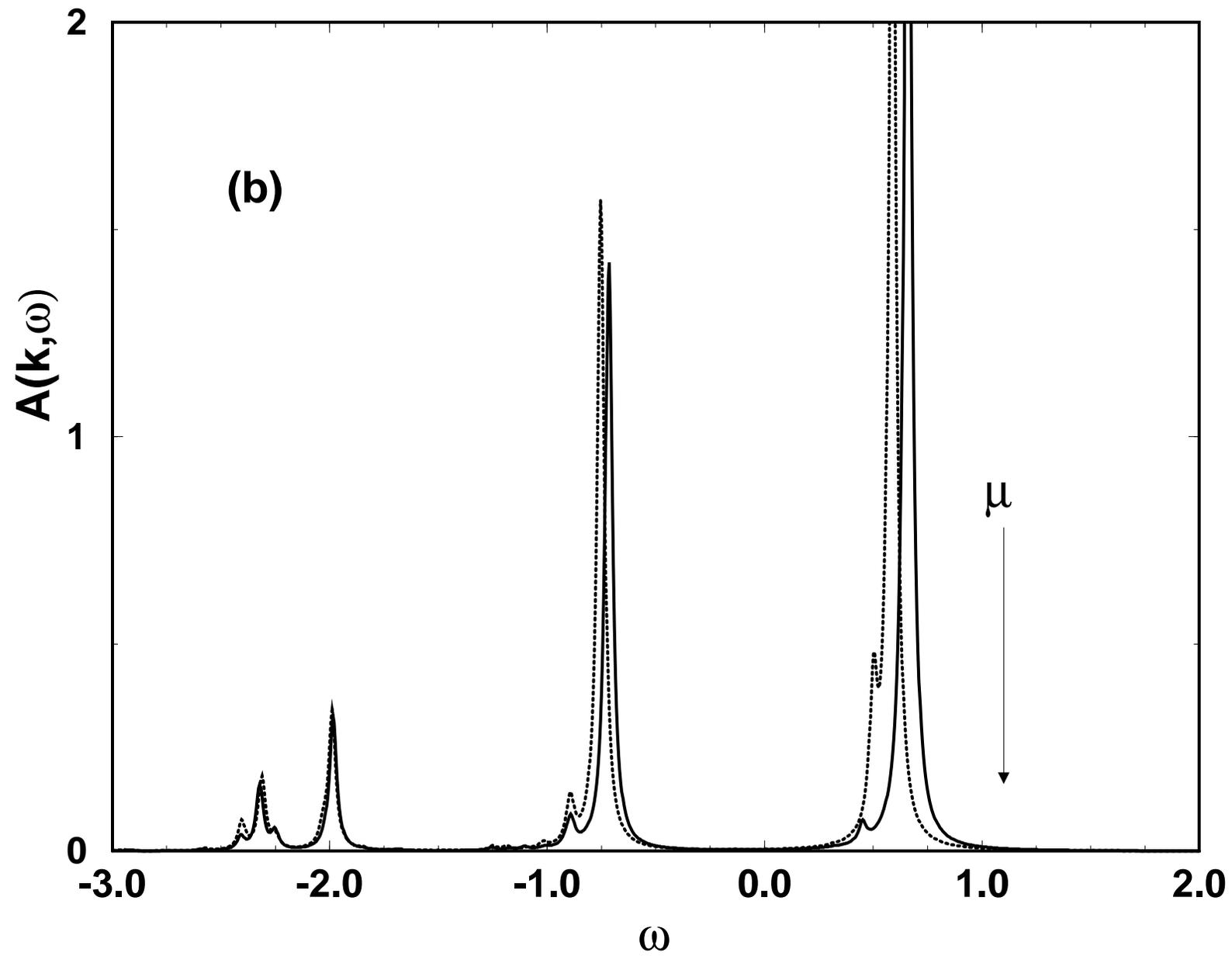
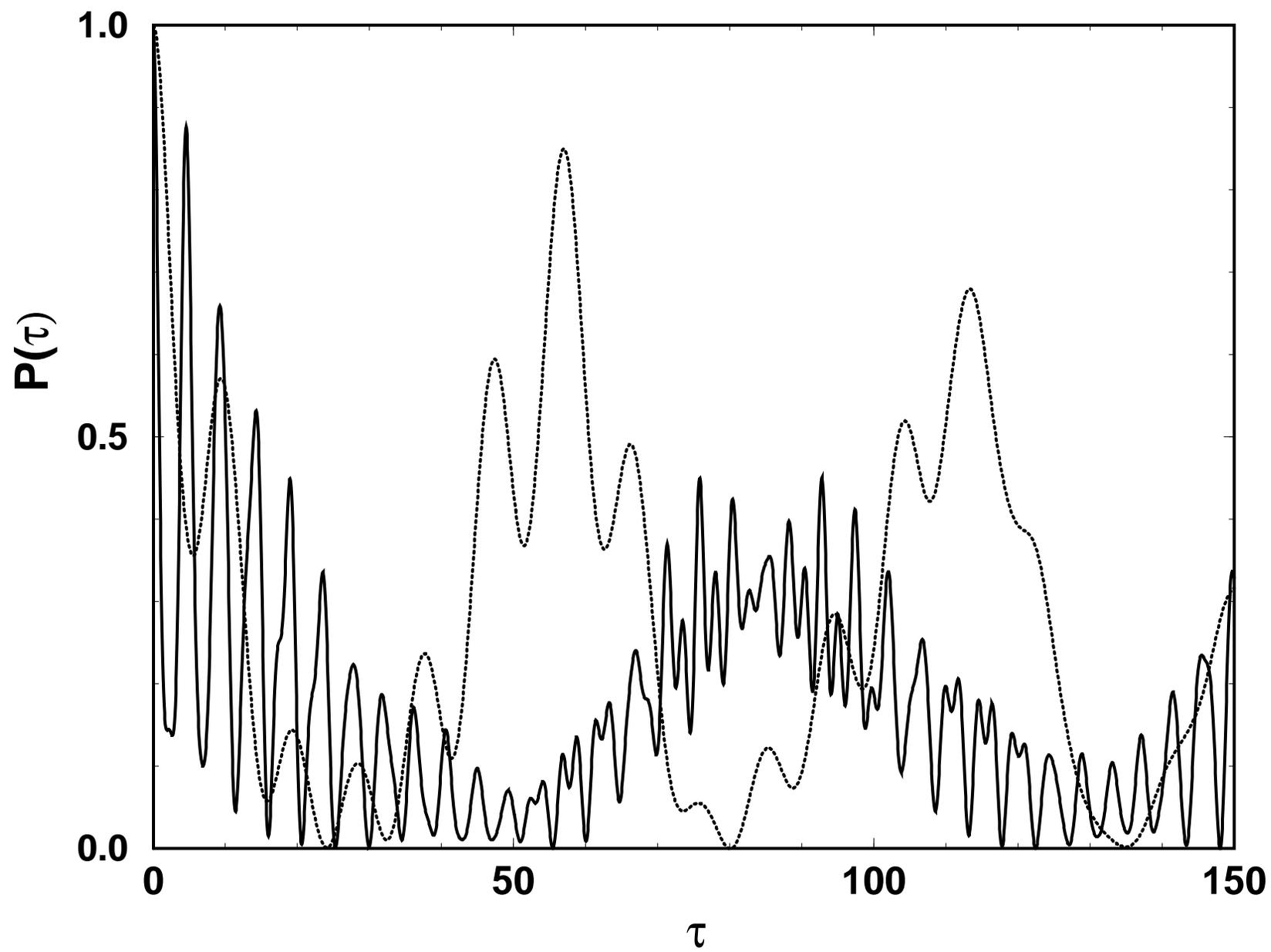
\end{document}